\documentclass{webofc}
\usepackage[varg]{txfonts}   
\usepackage{booktabs}

\begin{document}
\title{Reinterpretation of classic proton charge form factor measurements} 

\author{\firstname{M.} \lastname{Mihovilovi\v{c}}\inst{1,2,3}\fnsep\thanks{\email{miha.mihovilovic@ijs.si}} 
 \and \firstname{D.~W.} \lastname{Higinbotham}\inst{4}
 \and \firstname{M.} \lastname{Bevc}\inst{2}
 \and \firstname{S.} \lastname{\v{S}irca}\inst{1,2}
}

\institute{
Jo\v{z}ef~Stefan~Institute, SI-1000 Ljubljana, Slovenia
\and 
Faculty~of~Mathematics~and~Physics, University~of~Ljubljana, SI-1000 Ljubljana, Slovenia
\and
Institut~f\"{u}r~Kernphysik, Johannes~Gutenberg-Universit\"{a}t~Mainz, DE-55128~Mainz,~Germany
\and
Thomas Jefferson National Accelerator Facility, Newport News, VA 23606, USA
}

\abstract{
In 1963, a proton radius of $0.805(11)~\mathrm{fm}$ was extracted from electron scattering data and this classic value has been used in the standard dipole parameterization of the form factor.  In trying to reproduce this classic result, we discovered that there was a sign error in the original analysis and that the authors should have found a value of $0.851(19)~\mathrm{fm}$. We additionally made use of modern computing power to find a robust function for extracting the radius using this 1963 data's spacing and uncertainty. This optimal function, the Pad\'{e}~$(0,1)$ approximant, also gives a result which is consistent with the modern high precision proton radius extractions.   
}

\maketitle

\section{Introduction}
The proton charge radius, $r_\mathrm{E}$, is the conventional measure for the size of the proton, a fundamental constituent of matter. This constant is defined as the derivative of the proton charge form factor, $G_\mathrm{E}^p$, at zero four-momentum transfer, $Q^2=0$:
\begin{eqnarray}
     r_\mathrm{E}^2 \equiv -6\hbar^2 \left.\frac{dG_\mathrm{E}^p}{dQ^2}\right|_{Q^2 = 0}\,, \label{eq:1}
\end{eqnarray}
and can be determined by both hydrogen spectroscopy and elastic lepton scattering~\cite{Miller:2018ybm}.
The first determination of the radius was done with elastic
electron scattering data by Hand et al.~\citep{Hand}, who determined the radius of $0.805(11)\,\mathrm{fm}$, the value used in the standard dipole parameterization of the form factor~\citep{Hofstadter1958,Weisenpacher:2000ip}. The original study was followed by several decades of dedicated nuclear scattering and spectroscopic experiments, which led to a recommended value for the proton charge radius of 0.8791(79)~fm~(CODATA 2010,~\cite{Mohr:2012tt}). This result was called into question when the extremely precise spectroscopic measurements on muonic hydrogen~\cite{Pohl2010, Antognini2013} reported a significantly smaller value of $0.84087(39)~\mathrm{fm}$. The observed discrepancy, colloquially known as ``the proton radius puzzle''~\cite{Pohl:2013yb} motivated several new experiments~\cite{Beyer2017, Hessels2019, pRad2019, Fleurbaey:2018fih}. These experiments have been accompanied by different reanalyses of the existing data~\cite{Horbatsch:2015qda,Griffioen:2015hta,Higinbotham:2015rja,Lee:2015jqa,Graczyk:2014lba,Lorenz:2014vha,Horbatsch:2016ilr,Alarcon:2018zbz}, focusing on data of Bernauer~et~al.~\cite{Bernauer2010,Bernauer:2013tpr}. In this paper we follow a different path and revisit the first data of Hand et al., and evaluate their result by using modern analysis techniques.    

\section{The classical approach}
\label{sec:1}
In the first determination of the radius, existing data on proton charge form factor from five different measurements were considered~\cite{ Littauer:1961zz, Bumiller:1961zz, Drickey1962, Yount1962, Lehmann:1962dr}, as noted in Table~\ref{tab:data}. 
\begin{table}[!hb]
    \centering
    \caption{Summary of the experimental data considered in the analysis. For each data set, the columns represent the number of measured points, the minimal and maximal value of four-momentum transfer at which $G_\mathrm{E}^p(Q^2)$ was measured, and the average experimental uncertainty.\vspace*{1mm}}
    \begin{tabular}{r|cccc}
        \toprule
                 & number of & $Q^2_{\mathrm{min}}$ &  $Q^2_\mathrm{max}$ & average  \\[2mm]
         Authors & data points & $[\mathrm{fm}^{-2}]$ & $[\mathrm{fm}^{-2}]$ & uncertainty  \\
         \midrule
         Litauer~et~al.~\citep{Littauer:1961zz} & 4 & 2. & 8. & 0.251\\
         Bumiller~et~al.~\citep{Bumiller:1961zz} & 10 & 0.36 & 10. & 0.051 \\
         Drickey~et~al.~\citep{Drickey1962} & 4 & 0.3 & 2.2 & 0.006 \\
         Yount~et~al.~\citep{Yount1962}& 3 & 0.28 & 1.3 & 0.016 \\
         Lehmann~et~al.~\citep{Lehmann:1962dr} & 6 & 0.3 & 2.98 & 0.012 \\
    \end{tabular}
    \label{tab:data}
\end{table}

In an attempt to reconstruct the radius of $0.81\,\mathrm{fm}$ we followed the original analysis approach and compared the data to the quadratic function in $Q^2$:
\begin{eqnarray}
  G_{\mathrm{quadratic}}(Q^2) = 1 - \frac{r_\mathrm{E}^2}{6} Q^2 + a Q^4\,. \label{model:1}
\end{eqnarray}
This model depends on two free parameters: the radius, $r_\mathrm{E}$,  in front of the linear term, and the parameter $a$ that determines the curvature of the function. Since the data are normalized, the constant term of the model is simply $1$.  
In the first step the two parameters were determined by fitting Eq.~(\ref{model:1}) to the data with $Q^2 \leq 3\,\mathrm{fm^{-2}}$, considering the entire region with the high density of experimental points. The obtained results were $r_\mathrm{E} = 0.819(21)\,\mathrm{fm}$ and $a = 0.00787(309)\,\mathrm{fm^{4}}$. However, the radius obtained in this manner should not be trusted since the true shape of the $G_\mathrm{E}^p(Q^2)$ may be more complex than a second order polynomial. At $Q^2 \approx 3\,\mathrm{fm^{-2}}$ the contributions of the $Q^6$ and $Q^8$ terms are not negligible and their omission from the fit causes a systematic shift in the determined radius. 

\begin{figure}[t!]
\begin{center}
\includegraphics[width=0.7\linewidth]{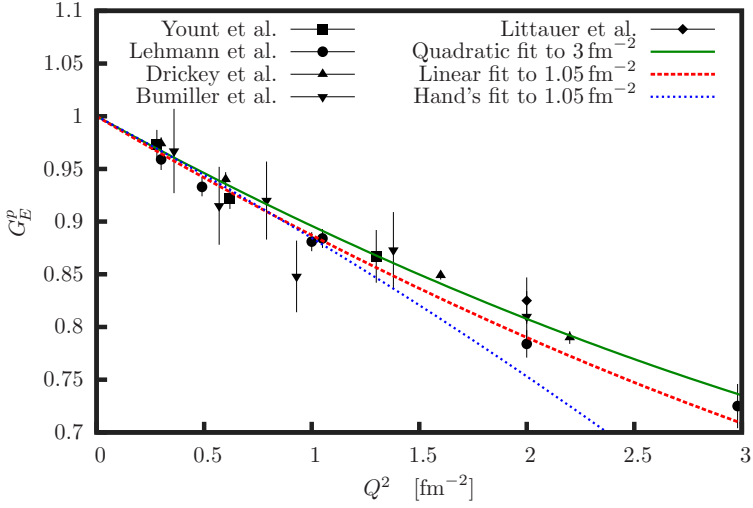}
\end{center}
\caption{The experimental data~\cite{ Littauer:1961zz, Bumiller:1961zz, Drickey1962, Yount1962, Lehmann:1962dr} considered in the analysis. The solid green line shows model~(\ref{model:1}) when both $r_{\mathrm{E}}$ and $a$ are fitted to the data with  $Q^2 \leq 3\,\mathrm{fm^{-2}}$. The dashed red line shows the results when $r_{\mathrm{E}}$ is fitted to the data with  $Q^2 \leq 1.05\,\mathrm{fm^{-2}}$, while the parameter $a = 0.00787\,\mathrm{fm^{4}}$ is kept constant. The blue dotted line corresponds to the original result of Hand~et~al., assuming $a = -0.00787\,\mathrm{fm^{4}}$. }\label{fig:1}
\end{figure}

To avoid model dependent bias in the radius extraction, the contributions of higher order terms should be kept minimal. The way Hand achieved this with a model, such as Eq.~(\ref{model:1}), is by keeping the parameter $a$ at a value determined in their first step and then only fitting the radius, using data with $Q^2 \leq 1.05\,\mathrm{fm^{-2}}$. Assuming that the determined value for $a$ is a good estimate for the size of the $Q^4$ term, this preserves the curvature of the model. Additionally, we were able to determine that at $1\,\mathrm{fm^{-2}}$ the $Q^4$ term contributes less than a percent to the value of $G_\mathrm{E}^p$. Hence, even a $10\,\mathrm{\%}$ error in the value of $a$ would result in a modification of the form-factor much smaller than the statistical uncertainty of each measurement. Hence, the described two step fitting technique should result in a more reliable estimate of the proton charge radius. We determined it to be $r_\mathrm{E} = 0.851(19)\,\mathrm{fm}$, which is inconsistent with the original result (see Fig.~\ref{fig:1}). The obtained value is $5\,\mathrm{\%}$ larger than the original radius while its uncertainty is almost twice as large as the uncertainty of the first result. 

\begin{figure}[htb]
\begin{center}
\includegraphics[width=0.9\linewidth]{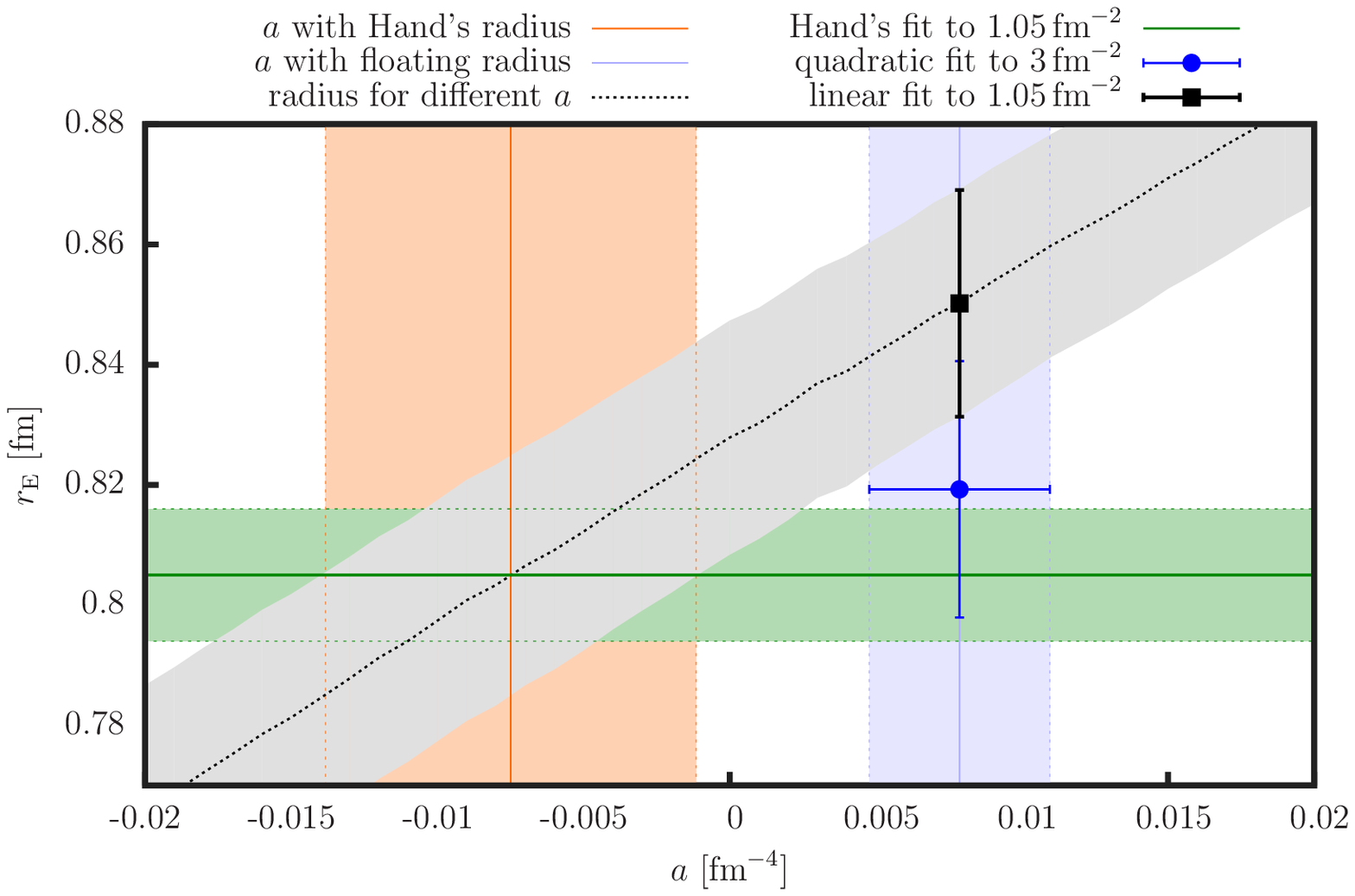}
\end{center}
\caption{The relation between parameters $r_\mathrm{E}$ and $a$ that determine the model~(\ref{model:1}). The green band denotes the original result of Hand~et~al.~\citep{Hand}. The blue point represents the result of the analysis when both parameters are free and the model is fitted to the data with $Q^2<3.00\,\mathrm{fm^{-2}}$. The vertical blue band indicates the value of the parameter $a$. The black point shows the final radius obtained by using the original two step approach of Hand~et~al. The gray line with the corresponding uncertainty shows how the extracted radius changes when $a$ is modified from $-0.02$ to $0.02$. The orange vertical band represents the result of the fit when only  $a$ is being fitted, while the radius is kept fixed at $0.805(11)\,\mathrm{fm}$. The cross-section of green, orange and gray bands defines the area of possible values of $a$ considered in the original analysis of Hand et al. The obtained result supports the hypothesis that a mistake has been made in the original analysis and that $a$ was considered with the wrong sign.  
}\label{fig:2}
\end{figure}

To find the source of the discrepancy the last step of the analysis was repeated with different values of $a$. Since $r_{\mathrm{E}}$ and $a$ are strongly correlated, it is important to evaluate the effect of $a$ on $r_{\mathrm{E}}$. Additionally, the original paper does not report the value of $a$. The analysis demonstrated in Fig.~\ref{fig:2} shows that the radius depends almost linearly on $a$ and reveals that the original value of $r_{\mathrm{E}}$ can be reproduced if $a$, determined in the first step of our analysis, is used, but with the opposite (wrong) sign.

To confirm this hypothesis, we again fitted model~(\ref{model:1}) to the data with $Q^2<1.05\,\mathrm{fm^{-2}}$, but this time kept the radius fixed at $0.805(11)\,\mathrm{fm}$ and adjusted only $a$. We obtained $a = -0.00749(63)\,\mathrm{fm^4}$, which strongly supports our assumption that a mistake was made in the original analysis. Additionally, our analysis has also revealed that the original study failed to acknowledge the uncertainty of $a$ in the determination of $r_\mathrm{E}$. Their analysis considered only statistical uncertainty and thus underestimated the final uncertainty of the radius. 

\begin{figure}[h!]
\begin{center}
\includegraphics[width=\linewidth]{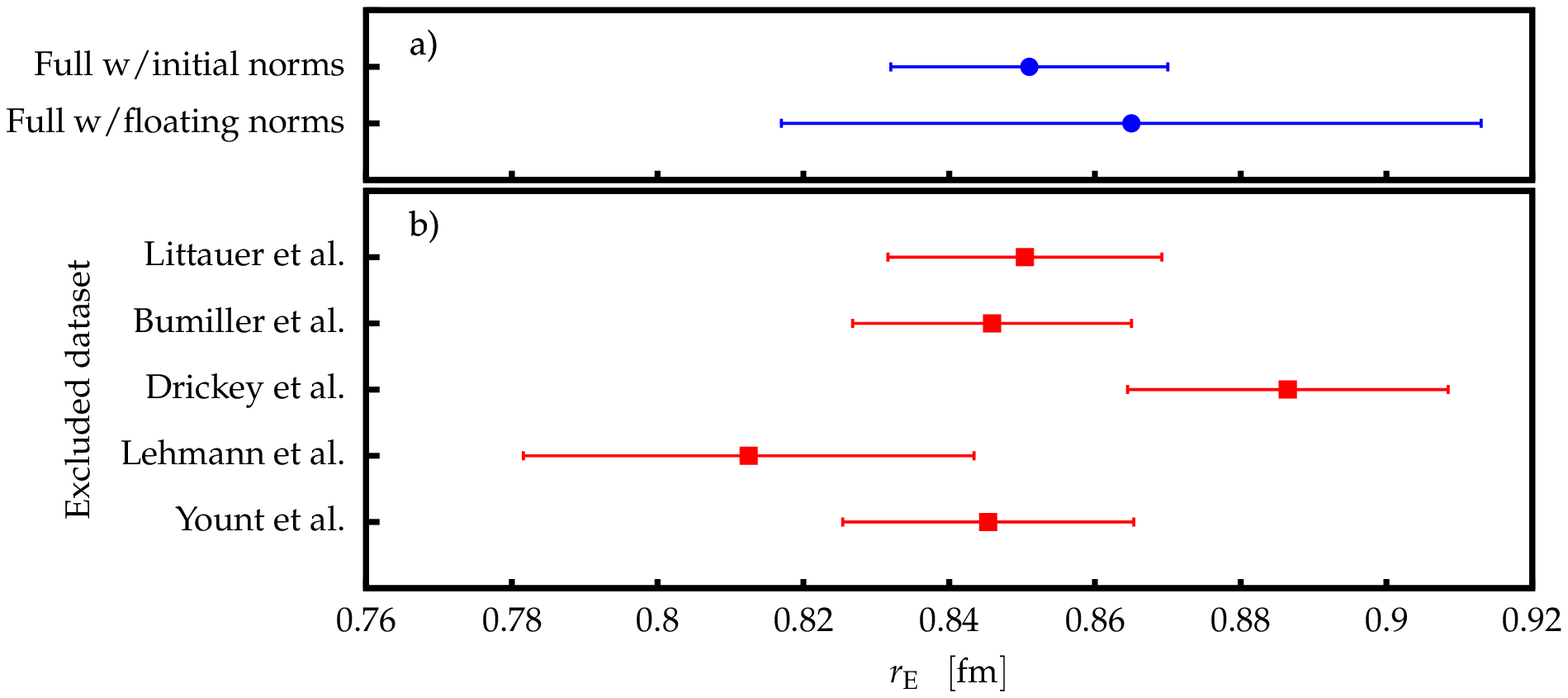}
\end{center}
\caption{The extracted values of the proton charge radius. {\bf a)} The difference between the value obtained with the fixed and floating normalization parameters. Addition of five free parameters significantly increases the uncertainty of the radius.  {\bf b)} Calculated radii when performing the analysis with only four out of five data sets, demonstrating a tension between the data sets of Drickey~et~al.~\cite{Drickey1962} and Lehmann~et~al.~\cite{Lehmann:1962dr}.}\label{fig:3}
\end{figure}

To test the stability of the extracted radius, we have repeated the analysis by using all combinations of four of the five data sets. The results presented in Fig.~\ref{fig:3} demonstrate the tension between the two most precise data sets, Drickey~et~al.~\cite{Drickey1962} and Lehmann~et~al.~\cite{Lehmann:1962dr}. The data of Lehmann~et~al. prefer a larger value of the proton charge radius and dominate the result when considering the data with small $Q^2$. The data of Drickey~et~al., on the other hand, favor a smaller proton charge radius and control the result at $Q^2 > 1.4\,\mathrm{fm^{-2}}$. While the discrepancy is too small to exclude a statistical fluctuation in the data, the most probable source of the tension are unaccounted for systematic effects, e.g. offsets in the absolute normalization of the reported data. The tension between the data is reduced if the normalizations of the data sets are kept as free parameters, as is being done in modern analyses of form factor measurements~\citep{Bernauer:2013tpr, Higinbotham:2015rja, Mihovilovic:2016rkr}, but does not disappear completely. Furthermore, introduction of additional five free parameters to the fits (normalizations) increases the variance of the extracted result and dilutes the significance of the extracted radius, which in the given case equals to $0.865(48)\,\mathrm{fm}$, see Fig.~\ref{fig:3}. 

\section{Robust analysis}
\label{sec:2}
The key problem of radius calculation is our ignorance of the true functional form of the proton charge form factor. Consequently, the form factor is approximated by various parameterizations. So far we considered function~(\ref{model:1}). Although the model was applied carefully to the data, it is not clear whether the quadratic function is an acceptable model for its description. The choice of a model can impact the result and can lead to a biased radius, i.e. a value that is systematically different from the true value. The bias is associated with the nature of the function and is typically smaller for functions with more free parameters. However, models with many parameters are justifiable only when data sets with large kinematic range and sufficient precision are available. Otherwise the variance of the radius increases to the level that the obtained result has no practical value. Hence, a model needs to be selected that exhibits a minimal bias of the extracted radius while keeping the variance of the result reasonably small. To achieve this, we have complemented the original analysis with a different technique based on a Monte-Carlo study of different form factor models, and are able to offer a more reliable determination of the radius. 



Since the majority of the available data were measured only at small $Q^2$ and with limited precision, we investigated only models that depend on up to three parameters in order to keep the uncertainty of the extracted radius below the difference between the two competing values of the proton radius problem. 
Beside model~(\ref{model:1}), we considered:
\begin{eqnarray}
     G_{\mathrm{cubic}} &=& 1 + n_1 Q^2 + n_2 Q^4 + n_3 Q^6\,, \label{model:2}\\
     G_{\mathrm{Pad\acute{e}~(0,1)}} &=& \frac{1}{1+ m_1 Q^2}\,,\label{model:3}\\
     G_{\mathrm{Pad\acute{e}~(0,2)}} &=& \frac{1}{1+ m_1 Q^2 + m_2 Q^4}\,,\label{model:4} \\
     G_{\mathrm{hybrid}} &=& \frac{1+n_1 Q^2 + n_2 Q^4}{1 + m_4 Q^8}\,, \label{model:5}\\
     G_{\mathrm{dipole}} &=& \frac{1}{(1+ m_1 Q^2)^2}\,,\label{model:6}
\end{eqnarray}
where $n_1$, $n_2$, $n_3$, $m_1$, $m_2$ and $m_4$ represent adjustable parameters of the models. Using these parameters the $r_\mathrm{E}$ for each model can be calculated using Eq.~(\ref{eq:1}). The quadratic (Eq.~(\ref{model:1})) and cubic functions (Eq.~(\ref{model:2})) were considered as well as four rational functions. They are interesting because, like the dipole model, they introduce higher order terms and define the curvature of the form factor at higher $Q^2$, although they depend on relatively few parameters. For completeness, we considered also the dipole model, which is known to report biased results~\citep{Bernauer:2016ziz}, but can serve as a test of our approach. 

\begin{figure}[ht!]
\begin{center}
\includegraphics[width=\linewidth]{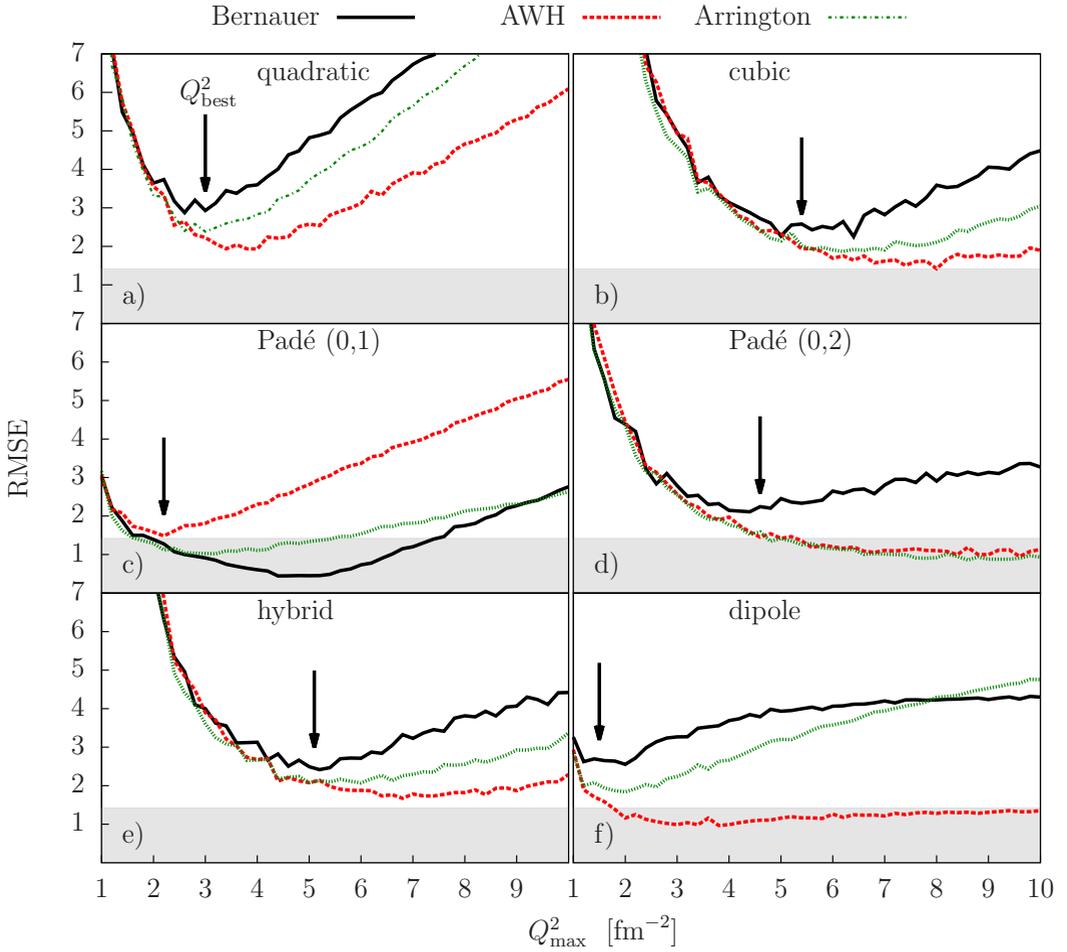}
\end{center}
\caption{Results of the Monte-Carlo study of the form-factor models~(\ref{model:1})~--~(\ref{model:6}). $\mathrm{RMSE}$ as a function of $Q^2_{\mathrm{max}}$, obtained with realistic form factor paramaterisations is used to evaluate the behaviour of each model. According to our selection criterion a model is appropriate for the analysis if the minima of all the curves on a given plot lie below the threshold of $\approx \sqrt{2}$. The selection threshold is marked on the plots with gray bands. The black arrows on each plot denote the positions of the highest minimum which
determines the interval $[0, Q^2_{\mathrm{best}}]$ of the data that should be considered in the fit.    
}\label{fig:4}
\end{figure}

The evaluation of the chosen models and tests of their capacity to reliably extract the radius can not be performed on the real data. Therefore we developed a Monte-Carlo simulation which generated many sets of pseudo data on a desirable kinematic interval using specific form factor models with known corresponding radii. These pseudo data were used to establish statistically relevant estimates on the size of the bias and variance of the extracted radius. The goal was to find a model that would (for a chosen kinematic range) return a radius with uncertainty smaller than $\sigma_{r_\mathrm{E}} \leq \sigma_0 = 0.02\,\mathrm{fm}$ and with the bias below $\Delta{r_\mathrm{E}}\leq 1/(2\sigma_0)$. Therefore, we have defined the estimator
\begin{equation}
\mathrm{RMSE} = \sqrt{\left(\frac{2 \Delta r_\mathrm{E}}{\sigma_0}\right)^2 + \left(\frac{\sigma_{r_\mathrm{E}}}{\sigma_0}\right)^2}\, \label{eq:2}
\end{equation}
which combines both conditions and could be used to quantify the quality of the selected model and search for the model with $\mathrm{RMSE}\leq \sqrt{2}$. The six models were tested by using the parameterization of Bernauer~et~al.~\citep{Bernauer:2013tpr} determined from real data, the fifth-order continued-fraction model of Arrington and Sick~\citep{Arrington2007}, and the theoretical prediction of Alarcon, Higinbotham, Weiss and Ye~\cite{Alarcon:2018zbz}. For each parameterization the pseudo data were generated and studied on the interval $[0, Q^2_\mathrm{max}]$. The results of the analysis are gathered in Table~\ref{tab:fits} and presented in Fig.~\ref{fig:4}. 

\begin{table}[!h]
    \centering
    \caption{Summary of the Monte-Carlo study of the form-factor models~(\ref{model:1})~--~(\ref{model:6}). For every model listed in column one, the table shows the results for the most pessimistic case, as can be seen in Fig.~\ref{fig:4}. Column two shows the ``best'' value of $Q^2_{\mathrm{max}}$ at which $\mathrm{RMSE}$ reaches its minimum and defines the range of the data $[0, Q^2_{\mathrm{best}}]$ to be used in the fit and in the extraction of the radius. Columns three and four contain the expected bias (extracted minus input radius) and uncertainty of the radius obtained with a chosen model.  The best $\mathrm{RMSE}$ values for a specific model are presented in column five. A threshold for a good model is arbitrarly set at $\sqrt{2}$, see column six. The last two columns show the values of the proton charge radius extracted from the data, together with their standard errors.\vspace*{1mm}}
    \begin{tabular}{r|ccccc|cc}
         \toprule
          & \multicolumn{5}{c}{SIMULATION} &  \multicolumn{2}{|c}{DATA}\\
         \midrule
         form factor      & $Q^2_{\mathrm{best}}$ & simulated   & simulated  & RMSE & acceptable &extracted  & standard   \\
         model &  & bias & uncertainty & & &radius & error  \\
         & $[\mathrm{fm}^{-2}]$ & $[\mathrm{fm}]$ & $[\mathrm{fm}]$ & & & $[\mathrm{fm}]$ & $[\mathrm{fm}]$\\ 
         \midrule
         quadratic & $2.9$ & $-0.023$ & $0.037$ & $2.93$ & no & $0.827$ & $0.023$ \\
         cubic & $5.4$ & $-0.016$ & $0.038$ & $2.52$ & no & $0.848$ & $0.032$ \\
         Pad\'{e} $(0,1)$ & $2.2$ & $\phantom{-}0.011$ & $0.022$ & $1.54$ & yes & $0.841$ & $0.009$ \\
         Pad\'{e} $(0,2)$ & $4.6$ & $-0.015$ & $0.028$ & $2.09$ & no & $0.826$ & $0.026$ \\
         hybrid & $5.1$ & $-0.016$ & $0.037$ & $2.49$ & no & $0.843$ & $0.032$ \\
         dipole & $1.5$ & $-0.022$ & $0.029$ & $2.63$ & no & $0.854$ & $0.019$ 
    \end{tabular}
    \label{tab:fits}
\end{table}

At small momentum transfers, the value of $\mathrm{RMSE}(Q^2_\mathrm{max})$ is governed by the variance, which decreases with the increasing number of data points considered in the fit. For large $Q^2_\mathrm{max}$, the model is no longer capable of satisfactorily describing the data. Consequently, the extracted radius  becomes biased and the $\mathrm{RMSE}(Q^2_\mathrm{max})$ again starts to increase. The position of the minimum determines the ideal momentum transfer range over which a given model gives the most reliable radius for a chosen form factor parameterization. Unfortunately, since we do not know the true functional form of the charge form factor, one cannot simply select a minimum from a single specific parameterization.  Thus, we try to be conservative and choose the minimum with the highest $\mathrm{RMSE}$ value, $Q^2_{\mathrm{best}}$, assuming that the form-factor parameterizations considered in the analysis form a representative set of functions and that the true form factor may be somewhere inbetween.  

\begin{figure}[h!]
\begin{center}
\includegraphics[width=\linewidth]{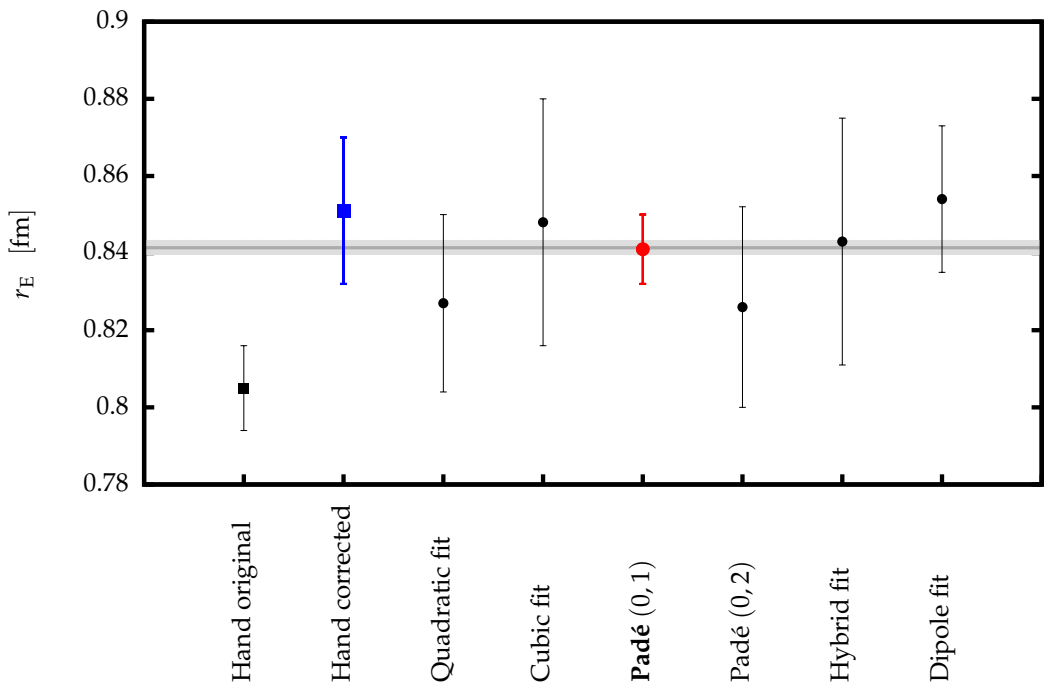}
\end{center}
\caption{The comparison of the extracted proton charge radii. The square points show the value calculated with the classical approach described in section~\ref{sec:1} and the original result of Hand~et~al.~\cite{Hand}. The circles represent the model-dependent extractions of the radius obtained with the new analysis technique presented in section~\ref{sec:2}.
The error bars show corresponding standard errors. According to the Monte-Carlo simulation the most robust estimate for the radius can be obtained using model~(\ref{model:3}), shown with the red circle. The gray band represents the new recommended value (CODATA 2018, ~\citep{CODATA2018}). }\label{fig:5}
\end{figure}

Once the $Q^2_{\mathrm{best}}$ for each of the models was estimated, the data could be fitted on the interval $[0,Q^2_{\mathrm{best}}]$ and the proton charge radius could be determined. The results of the fits to the real data are shown in Table~\ref{tab:fits}, Table~\ref{tab:higher} and in Fig.~\ref{fig:5}. However, the Monte-Carlo analysis demonstrates that only model~(\ref{model:3}) satisfies the condition for the $\mathrm{RMSE}=1.54 \approx \sqrt{2}$. All other models have $\mathrm{RMSE}$ values larger than $2$, which means that the radius results will not meet our criterion regarding the bias and variance. While quadratic and dipole functions are expected to have a large bias and should therefore be excluded, the remaining functions could still be considered, because their $\mathrm{RMSE}$ values are dominated by the large variance, but the calculated radii are expected to have large uncertainties. Hence, our best estimate for the radius is obtained with the Pad\'{e}~$(0,1)$ approximant, yielding the radius of $0.841(9)\,\mathrm{fm}$. 

\begin{table}[!hb]
    \caption{The parameters for the form-factor models (\ref{model:1}), (\ref{model:3}), (\ref{model:5}) and~(\ref{model:6}), which have more than one free parameter. Each table shows the values for a given model extracted from the data. The relative contributions of the terms equipped with the given parameters to the total value of the form-factor at $Q^2_{\mathrm{best}}$ are also presented. The alternating signs of the parameters of the quadratic model ($r, a$) and cubic function ($n_1, n_2, n_3$) indicate that the true nature of the form-factor is more complex than a low order polynomial, thus requiring higher-order terms to match its slope and the curvature in a chosen $Q^2$-range. The positive values of $m_1, m_2$ and $m_4$ show that the Pad\'{e}~$(0,2)$ and the hybrid model do not have poles, while automatically ensure a correct asymptotic behaviour of the form-factor. The large uncertainties of the higher-order terms $(n_2, n_3, m_2, m_4)$ are governed by the large uncertainties of the available measurements. 
    \vspace*{1mm}}
    \centering
\begin{minipage}{0.49\linewidth}

    \begin{tabular}{c|c|c}
         \toprule
         \multicolumn{3}{c}{qudratic} \\
         \midrule
         parameter & extracted & relative  \\
                   & value & significance\\
         \midrule
         $r$ & $0.827(23)\,\mathrm{fm}$ & $-1.30$\\
         $a$ & $6.0(24)\,\mathrm{fm}^4$ & $\phantom{-}0.30$\\
    \end{tabular}
    \vspace*{5mm}
\end{minipage}
\begin{minipage}{0.49\linewidth}
    \begin{tabular}{c|c|c}
         \toprule
         \multicolumn{3}{c}{Pad\'{e} $(0,2)$} \\
         \midrule
         parameter & extracted & relative  \\
                   & value & significance\\
         \midrule
         $m_1$ & $2.92(18)\,\mathrm{fm}^2$ & $-0.94$\\
         $m_2$ & $1.7(25)\,\mathrm{fm}^4$ & $-0.14$\\
    \end{tabular}
    \vspace*{5mm}    
\end{minipage}
\begin{minipage}{0.49\linewidth}
    \begin{tabular}{c|c|c}
         \toprule
         \multicolumn{3}{c}{cubic} \\
         \midrule
         parameter & extracted & relative  \\
                   & value & significance\\
         \midrule
         $n_1$ & $-3.08(24)\,\mathrm{fm}^2$ & $-1.24$\\
         $n_2$ & $\phantom{-}11.3(57)\,\mathrm{fm}^4$ & $\phantom{-}0.95$\\
         $n_3$ & $-40.2(322)\,\mathrm{fm}^6$ & $-0.71$\\
    \end{tabular}
\end{minipage}
\begin{minipage}{0.49\linewidth}
    \begin{tabular}{c|c|c}
         \toprule
         \multicolumn{3}{c}{hybrid} \\
         \midrule
         parameter & extracted & relative  \\
                   & value & significance\\
         \midrule
         $n_1$ & $-3.04(22)\,\mathrm{fm}^2$ & $-1.27$\\
         $n_2$ & $\phantom{-}8.9(40)\,\mathrm{fm}^4$ & $\phantom{-}0.74$\\
         $m_4$ & $ 275(236)\,\mathrm{fm}^8$ & $-0.63$\\
    \end{tabular}
\end{minipage}
    \label{tab:higher}
\end{table}

\section{Conclusions}

\begin{figure}[ht]
\begin{center}
 \includegraphics[width=1\textwidth]{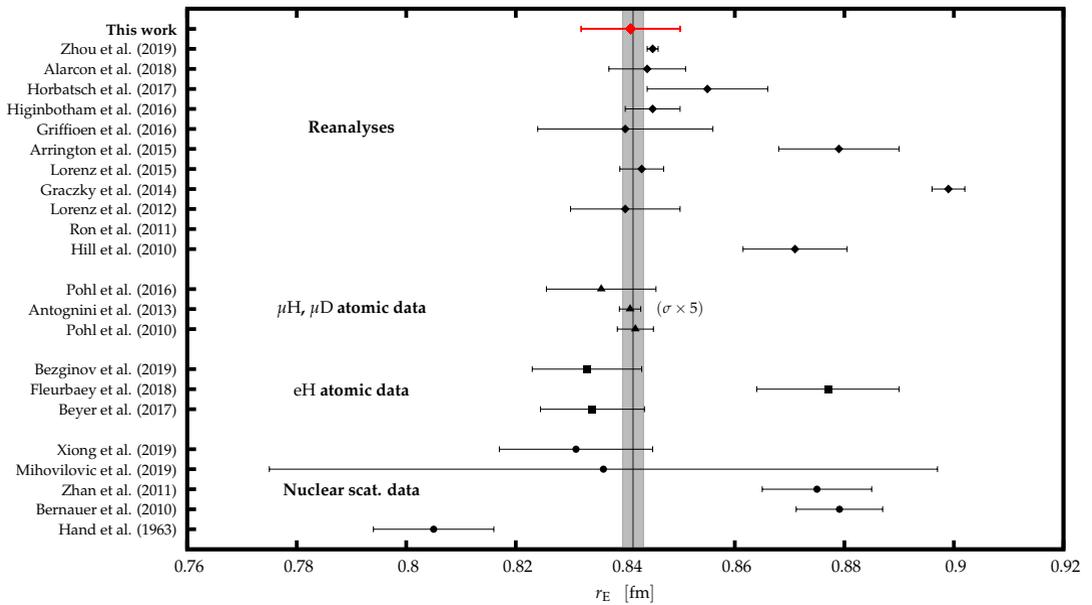}
 \caption{ 
 The result of this work compared to other extractions of the proton charge radius. Full circles show findings of modern nuclear scattering experiments (\citep{ pRad2019, Bernauer:2010zga, Zhan:2011ji, Mihovilovic:2019vkh}) together with the original result of Hand~et~al~\citep{Hand}. Full squares represent values obtained from the recent atomic hydrogen spectroscopy measurements (\citep{Beyer2017,Hessels2019,Fleurbaey:2018fih}). The triangles denote values determined from the muonic hydrogen (deuterium) measurements (\citep{Pohl:2010zza, Antognini:1900ns, Pohl:2016tqq}). The uncertainties of data from Pohl~et~al. and Antognini~et~al. are multiplied by factor $5$ for clarity. The diamonds show recent reanalyses of the electron scattering experiments (\citep{Griffioen:2015hta, Higinbotham:2015rja, Graczyk:2014lba,  Lorenz:2014vha,Horbatsch:2016ilr, Alarcon:2018zbz, Arrington:2015ria, Hill:2010yb, Ron:2011rd, Lorenz:2012tm,  Zhou:2018bon}). The gray line with the corresponding band is the recommended value (CODATA 2018,~\citep{CODATA2018}).
 \label{fig_World}}
 \end{center}
\end{figure}

In this paper we reanalyzed the proton charge form factor data from classical experiments performed in the 1960s by utilizing modern analysis tools that were not available at the time of the original analysis. Repeating the steps of Hand et al., we determined the radius to be $0.851(19)\,\mathrm{fm}$, a value which is $5\,\mathrm{\%}$ larger than the result of the original paper. Using Monte-Carlo simulation we determined that the observed discrepancy is most probably related to a mistake in the interpretation of the $Q^4$-term when fitting the radius. To evaluate and minimize the dependence of the radius on the model applied in the analysis, the classical approach was superseded by a Monte Carlo-based analysis using pseudo-data generated with realistic form-factor parameterizations. In this approach the most appropriate fitting interval and the model function was selected by using a predefined selection criterion $\mathrm{RMSE}\leq \sqrt{2}$. Among the considered functions only Pad\'{e}~$(0,1)$ fulfilled the set condition. Using this function the best estimate for the proton charge radius was determined to be $0.841(9)\,\mathrm{fm}$. The obtained result is in good agreement with recent extractions of the radius and with the new recommended value (CODATA2018,~\citep{CODATA2018}), see Fig.~\ref{fig_World}.  Minimization of the model dependence of the extracted radius is key for reaching consistent interpretation of the modern electron scattering data. Here we offer an approach, which, relying on  predefined selection criterion and using Monte-Carlo simulations, simultaneously examines both the model bias and variance. The method successfully applied to the data of Hand~et~al. can be directly extended to more complex models and used for a robust interpretation of the recent data.  

\section*{Acknowledgements}
This work is supported by the Federal State of Rhineland-Palatinate, by the Deutsche Forschungsgemeinschaft with the Collaborative Research Center 1044, by the Slovenian Research Agency under Grant P1-0102, and by 
Jefferson Science Associates which operates Jefferson Lab for the U.S. Department of Energy under contract DE-AC05-06OR23177.
\bibliographystyle{frontiersinHLTH&FPHY} 
\bibliography{ProtonRadius}
\end{document}